# Organizing Linked Data Quality Related Methods

Philippe A. MARTIN
ESIROI I.T., EA2525 LIM, Uni. of La Réunion, Sainte Clotilde, France
(+ adjunct researcher of the School of ICT, Griffith Uni., Australia)

**Abstract** - *This article presents the top-level of an ontology categorizing and generalizing best practices and quality criteria or measures for Linked Data. It permits to compare these techniques and have a synthetic organized view of what can or should be done for knowledge sharing purposes. This ontology is part of a general knowledge base that can be accessed and complemented by any Web user. Thus, it can be seen as a cooperatively built library for the above cited elements. Since they permit to evaluate information objects and create better ones, these elements also permit knowledge-based tools and techniques – as well as knowledge providers – to be evaluated and categorized based on their input/output information objects. One top-level distinction permitting to organize this ontology is the one between content, medium and containers of descriptions. Various structural, ontological, syntactical and lexical distinctions are then used.*

## 1 Introduction

How should data or knowledge be represented and published so it can most easily be retrieved, re-used and managed? Then, how to compare or evaluate knowledge statements, knowledge bases (KBs), knowledge management techniques, KB management systems (KBMSs) and knowledge providers? Many complementary or alternative "knowledge sharing supporting *elements*" – and kinds of elements – have been proposed to provide partial answers to these research questions. For example: Semantic Web approaches [1-2], knowledge sharing languages (e.g., those provided by the W3C and more general ones [3-6]), ontologies [7-13], methodologies [14-15], best practices or design patterns [16-18], categories of evaluation criteria or measures [19-22], knowledge quality evaluation queries [23-25], benchmarks [26-27], techniques [28-29], software, etc. (these references are given to illustrate each "kind of element"; many of them will also later be referred to in this article; giving them *now* permits to group them by "kind" in the "references" section of this article).

However, it is difficult for a knowledge provider or a KBMS developer to know about all these elements or their sub-elements, to compare them, choose between them, combine them and have a *synthetic organized view* of what can or should be done for knowledge sharing purposes. Indeed, these elements often do not use similar terminologies or categorizations, and *no ontology or library* has been proposed *to compare, index, organize and generalize* these elements (which are at various levels of abstraction and may be contradictory). *The top-level ontology presented in this article is a step in that direction. Thus, the goal of this article is to show how the problem cited at the beginning of this paragraph can be addressed.* This article refers to this ontology or library as "this knowledge criteria/quality ontology" or simply "this (top-level) ontology". It is currently focused on elements related to Linked Data [2].

As with any other ontology, the bigger *and more organized* it will become, the more useful it will be for the above cited knowledge sharing and retrieval tasks. (For the knowledge operationalization tasks, bigger is no longer better but eases the selection of knowledge for modules of relevant sizes and content for an application; this article and its top-level ontology only address knowledge sharing and retrieval tasks). This ontology is part of a KB published on-line via the WebKB knowledge server (usable at http://www.webkb.org). This KB – and hence the ontology presented in this article – can be cooperatively extended by any Web user via this server. To enable this, WebKB uses an abstract model and editing protocols [29] allowing its KB to be *consistent* and organized even though knowledge statements come from different sources and hence can *contradict* each other. Because of space restrictions this model is only quickly introduced in this article (it is not described) but this knowledge quality ontology exploits it and categorizes some of its features.

This ontology is only about information objects. Indeed, once information objects can be evaluated, knowledge management tools and techniques can be compared or evaluated with respect to (w.r.t.) the qualities of the information objects that they allow as input and output, or lead their users to produce. Similarly, knowledge providers can be evaluated w.r.t. the information objects they have provided. Thus, unlike the "Semantic Web (SW) Topics Ontology" [10], this ontology does not attempt to semantically organize – i.e., does not attempt to use specialization relations, "part of" relations or other relations to organize – knowledge management tools, techniques or processes (e.g., tools and techniques for knowledge extraction, retrieval, matching, merging, representation, inferencing, validation, edition, annotation, modularization and publishing). This would be a huge task and, as for example illustrated by the "SW Topics Ontology", these processes are so intertwined that they are difficult to distinguish and organize in a *scalable* way, i.e., in a systematic and non-arbitrary way within a specialization hierarchy and a part-of hierarchy. Here, "non-arbitrary" implies the use of conceptual distinctions – and especially, partitions – that are clear enough to lead different persons to categorize a same thing at a same place in a specialization/part-of hierarchy (note: a hierarchy does not have to be a tree). Such distinctions and hierarchies significantly reduce implicit redundancies [14].

The uppermost conceptual distinction used by this ontology to permit non-arbitrary categorizations of information objects is the clear partition of information objects into either description-content, description-medium or description-container objects. In this article, *description-content* objects are conceptual categories, as well as formal/informal terms or statements referring to or defining these categories. They are interpretations or abstractions of a (real or imaginary) situation or object. E.g.: abstract models, ontologies, terminologies, languages and any of their sub-elements (e.g., the concept/relation types of RDF and OWL). D*escription-medium* objects are concrete model objects permitting to visually/orally/... present description-content objects. E.g.: graphical interface objects and syntax/style objects such as those specified by XML, CSS and XSLT. *Description-container* objects are the other information objects, i.e., non-physical objects permitting to store and manage description-content and description-medium objects. E.g.: files, file repositories, distributed databases and file servers.

The sections 2, 3 and 4 are respectively about the evaluation of description-content, description-medium and description-container objects. These sections relate, organize and generalize knowledge sharing best practices and quality criteria/measures from various sources. Some categories from each of the above referred articles are included in this quality ontology. [17-22] include the most complete lists of high-level categories that seemed to exist so far for Linked Data. All their categories are integrated in the quality ontology. The top-level of this ontology may be seen as validated by the fact it correctly organizes and generalizes all the quality-related categories that the author has found so far (this is a kind of "validation by usage"). On the other hand, this was obvious given the way this ontology was designed. The difficult part was to find this design. This ontology may also be seen as validated by the fact it follows the strongest best practices it categorizes (the ones that include or imply the weaker or more elementary best practices, e.g., the fact that a KB should at least be consistent). In [17-22], elements are only slightly categorized. This is shown by Section 5 which gives the four most organized quality related categorizations that seemed to exist for Linked Data so far. These categorizations are essentially only two levels deep and not always intuitive. Additional conceptual distinctions would be interesting, especially if they are "non-arbitrary" (under this condition, the more categories, the better).

## 2 Description content quality

This article cannot present the whole "quality ontology". It can only show its top-level and its principles, i.e., the way this ontology manages to organize the *main kinds* of methodological elements, best practices, quality characteristics (e.g., evaluation criteria, quality dimensions, the "data quality indicators" of [22], ...) and quality measures (e.g., the "scoring functions" and "assessment metrics" of [22], ...) that have been proposed for knowledge sharing purposes. This article only shows important elements of a *subtype hierarchy of quality measuring functions* on information objects, with the function result being a value (typically, numerical or boolean). Indeed, all quality related categories and subtype hierarchies can be automatically *derived* from the above subtype hierarchy. E.g., *relations* can be derived from boolean functions (Figures 3 includes an example) and, from the above subtype hierarchy, it is possible to derive the one for *quality characteristics* and the one for *"statements that have a certain (kind of) quality measure"* (alias, "statements that follow a certain (kind of) best practice").

There are many ways to categorize quality evaluations, e.g., according to the kind of objects they evaluate, and whether or not they take into account certain lexical, structural or semantic best practices. Figure 1 shows one intuitive uppermost categorization. In this article, indented lists show subtype hierarchies. In all such indented lists below, the XML namespace shortcut can be used but the prefix "pm:" is left implicit. "LDpattern:" is for [18], "LD:" for [20], "SF:" for [21] and "PD:" for [22-23]. C++/Java-like comments are used. Relation identifiers use nominal expressions and follow the common "graph reading convention", i.e., the last argument is the destination of the relation. Thus, a binary relation R(X,Y) can be read "<X> has for <R> <Y>" (notes: <X> and <Y> may include quantifiers; furthermore, a relation "X has for subtype Y" may also be read "any instance of Y is also an instance of X"). For functional relations, the last argument is the function result.

Figure 2 gives subtypes to pm:description_content_quality. Two points explain those subtypes. First, one handy partition for description-content_semantic-quality functions is the distinction between those that give "correctness" values for the evaluated object and those checking that it includes certain things. Second, for each kind of evaluated source object, there are various ways to categorize  i) functions that evaluate certain aspects of this kind of objects (e.g., the "correctness" and "conformity" aspects), ii) functions that evaluate "*related* objects", and iii) functions that differently aggregate the values returned by these functions. In this article, within the names of the last two kinds of functions, "description" is abbreviated by "descr" in order to make the hierarchy more readable. The adjective "*related*" refers to *actual or potential/allowed* relations. E.g., RDF (a description content) *allows* various kinds of textual or graphical notations (description media) – some being standards, some not – even if most RDF-based tools (description containers) only work with RDF/XML. Some evaluation functions may for example *better* rate an RDF-based tool that can handle *more* notations (e.g., by calling external translation tools).

```
quality   //function on an object with possibly other arguments;
          //this function returns a value (numerical, boolean, ...)
  content_based_quality   //at least based on the object content
  meta-statement_based_quality   //ie., on meta-statements on the object
    rating_based_quality   //at least based on ratings
```

Figure 1.  Important top types to organize quality measuring functions

```
description-content_quality   //subtype of above function pm:quality
  correctness   //one main kind of description-content_quality;
                // Figure 3 gives some s important subtypes
  conformity   //another main kind; Figure 5 gives important subtypes
  quality_of_this_descr_content   //to evaluate the source object
                                  // on all its criteria
    descr_content_quality_of_this_descr_content   //content-related
                                                  // aggregations
    quality_of_descr_media_related_to_this_descr_content
    quality_of_descr_containers_related_to_this_descr_content
```

Figure 2.  Important ways to evaluate description content quality

```
correctness   //of the evaluated object (statement or term
              // referring to a statement)
  LD:accuracy   //factual correctness of a statement (which
              // should be a belief) w.r.t. the world
  consistency   //reports all or some inconsistencies and
              // implicit contradictions
    consistency_of_this_statement_wrt_this_one  (ST,ST -> boolean)
              //this signature states that this function is boolean
              // and has exactly 2 statements as arguments;
              //one relation derivable from this function is:
              // pm:statement_consistent_with_this_one (ST,ST)
    consistency_and_non-redundancy_of_this_statement_wrt_this_one
              //signatures are inherited when there is no ambiguity
    consistency_of_this_KB (ST -> boolean)
      consistency_of_the_RDF_KB  //the KB must be an RDF KB
        consistency_of_SKOS_relations  //the measures of [24] are
              // subtypes of this type
        consistency_of_a_RDF_KB_tested_via_a_SPARQL_query
              // as in [23] and [25]
    LD:internal_consistency_fct  //SF:consistency_fct seems to
              // be an alias of this type
    LD:modeling_correctness_fct   //tests the correctness of the
              // "logical structure of the data"; LD and SF do not
              // precise if these last 2 types are dimensions or
              // functions; this is why "_fct" is a suffix here;
              // other relations and dimensions can be derived
    substatements_of_this_1st_statement_that_are_inconsistent_\
with_this_2nd_statement  (ST,ST -> set)
  consistency_ratio  //no restriction on the arguments but the
              // result is a number (a ratio)
    consistency_ratio_of_such_a_statement_in_this_statement
        (ST,ST -> number)  //"such a statement": the 1st argument
    consistency_ratio_of_all_substatements_in_this_statement
        (ST -> number)  //contextualized substatements
    consistency_ratio_of_this_KB (ST -> number)
    consistency_ratio_of_relations_on_this_term_in_this_statement
        (term -> number)
    consistency_ratio_of_this_relation_on_this_term_in_this_statement
        (ST,term -> number)
    PD:consistency  //"number of non-conflicting frames"
              // divided by the "number of frames"
```

Figure 3. Important types of functions to evaluate correctness

```
(pm%def_fct  ;;one of the operators for defining term, here a function
  pm%consistency_ratio_of_such_a_statement_in_this_statement
    (?s1 ?s2)  ;; parameters;  "?" is a KIF prefix for variable names;
              ;;as in all evaluation functions, the first argument
              ;; is the evaluated object;  ?s2 may be a whole KB
    (div (pm%cardinality ;;size of the set returned by "setofall"
          (setofall ?s (and  (pm%substatement ?s2 ?s)  (=> ?s ?s1)
                             (not (=> ?s2 (not ?s))) )))
         (pm%cardinality (pm%substatements ?s2)) ))
```

Figure 4. Full definition of one of the above functions, in KIF

[23] and [25] implement some quality measures (for whole RDF-based KBs) via SPARQL queries and SPIN rules. As noted in Figure 3, some of these measures are about consistency. They permit to check some aspects that KB building tools already verify or some complementary aspects. More classic and powerful consistency checking or other quality measurement cannot be done – or would be too complex and long to be done – via SPARQL queries. They require inference engines and definitions in expressive languages such as KIF [5]. The Common Logic ISO standard and its CLIF notation [6] are adaptations of the model and syntax of KIF but without features for specifying meta-statements, definitions and (monotonic or not) inference rules. This is why the data model (alias, meta-ontology) of WebKB is defined in KIF and why some definitions of its quality related ontology are in KIF. As illustrated by Figure 4, the functions referred to in this article are relatively easy to write in KIF. This figure shows the use of "pm%def_fct", one of the primitive definition operators of the data model of WebKB, fully defined in KIF. These operators also permit to check and handle lexical/namespace and semantic contextualizations in a KB. "%" is used instead of ":" for XML-like namespace shortcuts because "%" and "#" have other meanings in KIF. "Semantic contextualization" refers to the embedding of a statement within a meta-statement for specifying information without which the statement may be false, e.g., information about the time, place and source of the content of the statement. Every statement must have at least one source (author, source document, etc.). This permits to consider "every assertion that is not a definition" as a belief. This is the main reason the KB can stay consistent while still allowing contradictory beliefs.

Figure 3 gives some specializations for the first subtype of pm:description_content_quality. So far, all current quality measures related to Linked Data seemed to use the whole KB (data set) as implicit argument. They may thus be simpler to call or to understand but this is a loss of generality. Furthermore, most of these measures only work on "frames" ("objects" in object-oriented approaches), i.e, on the set of relations from a term. They do not work on any kind of statement. The hierarchy in Figure 3 shows how different but related evaluation functions and relations can be organized and generalized. Concept types can be derived too. An important one is pm:Statement_consistent_and_non-redundant_with_any_other_one_in_the_KB. Indeed, if a KBMS checks that each statement is of this type before allowing its insertion it into the KB, every statement has a place in the specialization hierarchy of this KB and can thus be automatically compared to other statements. This is enforced by the KB editing/sharing protocols of WebKB.

The KB editing/sharing protocols of WebKB also enforce the statement of relations – such a *pm:corrective_refinement* – between contradictory beliefs. This avoids implicit partial/total redundancies and permits people or applications to choose between these contradictory beliefs when choices have to be made. Thus, no arbitrary selections has to be made *a priori* by the KB owners. E.g., a user may specify that "when browsing the KB or within the results of his knowledge extraction queries" he wants to see only the most specialized *corrections* from certain kinds of sources. Furthermore, WebKB allows "default beliefs" and provides some default beliefs. E.g., a default belief for any user in WebKB is that, among contradictory statements, he believes the ones that *correct* the others without being contradictory between themselves. Since quality related measures can be specified as definitions or as beliefs, by creating or overriding some defaults beliefs, a user can easily specify which measures he believes in and wants to use, and how they should be combined. For more details on semantic/lexical contextualizations, default beliefs, and how to combine them, see the KIF formalization of the WebKB data model at http://www.webkb.org/kb/it/KSmodel.html

```
conformity  //reports on the existence/number of certain things or
            //patterns (thus, even SF:amount_of_data is a subtype of it)
  conformity_of_this_statement_wrt_this_requirement
       (ST,ST -> boolean)
  ratio_of_conformity_to_this_requirement_in_this_statement
       (ST,ST -> number)
  ratio_of_conformity_of_the_KB   //no argument restriction here
     LD:modeling_granularity (-> number)  //no argument
     PD:structuredness  //e.g., PD:coverage (number of objects
                        // with all relations of a schema) and
                        // PD:coherence (average of coverage for all terms)
     PD:completeness  //alias, LD:completeness (do all required
                      // terms/relations exist?)
        PD:intensional_completeness  //ratio (percentage) of
                                     // required relations in the KB
        PD:extensional completeness  //ratio of required terms
        PD:LDS_Completeness //ratio of terms with a given relation
     PD:relevancy  //alias LD:Boundedness, ratio of data relevant
                   // for an application
     PD:verifiability  //existence of information to check for
        //correctness; examples of subtypes: PD:traceability,
        // PD:provability,  PD:accountability; the following two
        // best practices are related to these subtypes:
        // - "providing another KB for tools that cannot perform
        //    complex  inferences" (LDpattern:materializing_inferences)
        // - "transforming the KB to conform to some models"
        //   (LDpattern:transformation_query)
     SF:validity  //no syntax errors, …; PD:verifiability is very
                  // related; PD:validity is a subtype
  representation_quality
     organization  //for formal and informal objects in the KB
                   //WebKB permits to organize both
        at_least_minimal_organization  //as defined and enforced in
                                       //WebKB; many other subtypes can be defined
        reachability  //PD:reachability when the object is a KB
           out-relations  //from the object; for a whole KB:
                          // PD:external_links, PD:outdegree, …
                          //the more out-relations, the better: this is the
                          // 4th basic rule for Linked Data [16]; the more widely
                          // known/deployed the target objects, the better
           in-relations  //to the object; for a KB: PD:indegree, …
        non-redundancy  //e.g., PD:intensional_conciseness, …
     expressiveness_economy  //avoidance of expressive constructs
                             // when this does not bias knowledge representation and
                             // reduce knowledge matching/inferencing/readability
     modeling_uniformity  //e.g., checks some lexical,structural
                          // or ontological conventions
        LD:directionality  //checks the consistency in the
                           // direction of relations
        use_of_the_graph-oriented_reading_convention //as for the
                                       //  5 above types, it is important for readability
        conformity_to_an_abstract_model_or_ontology_or_methodology
           conform_to_Ontoclean  //checks that the object (or each of
               // its sub-objects) is instance of at least one of the Ontoclean
               // 2nd-order types: (semi/anti/totally_)rigid_thing, etc.
           use_of_a_standard_model  //3rd basic rule for Linked Data
                                    // but for abstract models only
     quality_of_the_representation_of_terms  //see Figure 6
```

Figure 5. Important subtypes of functions to evaluate conformity

Figure 5 – and, its continuation, Figure 6 – show subtype relations between types of functions checking that within an object certain elements exist and are conform to a certain pattern. The various subtypes are semantically close. The first listed subtype can be re-used to write the other ones. This subtype hierarchy shows that the "current categorizations for Linked Data quality criteria and measures" (LD, PD, SF, ...) only cover particular cases. Thus, the current implementations of (some of) these measures also only cover particular cases. To save space, there is no repetition of types in this hierarchy (this applies to the next hierarchies too). Some of the types could clearly also appear at other places. The comments give some explanations for each of the types. The ones in bold and/or italics are the most important for categorization or re-use purposes.

```
  quality_of_the_representation_of_terms  //as in end of Figure 5
     identification_by_properly_formed_URIs  //checks that
             // objects are identified by HTTP URIs that can be
             // dereferenced by agents to find further information;
             // these are first two rules of Linked Data [16];
             // [18] gives some specializations to these best practices
     following_of_naming_conventions  //use of nouns, of a
                                      // loss-less naming style, …
     LD:referential_correspondence  //consistency and
                                    // non-redundancy of identifiers
     LD:typing  //checks that nodes are first-order typed
                // entities, not just strings, hence checks the
                // "Link Not Label" best practice [18]
     PD:vocabulary_understandability  //checks that terms have
                                      // human readable labels, …
        LD:intelligibility  //alias SF:comprehensibility? These two
                            // types appear to be only about terms
        PD:internationalization_understandability  //checks that
                                      // the language is specified
  quality_of_existing_or_derivable_relations
     use_of_binary_relations_only  //since this helps knowledge
                                   // matching and precision
     quality_of_existing_or_derivable_meta-statements  //and hence
                                   // relations from statements
        quality_of_existing_or_derivable_contexts //temporal ones,
                                   // spatial ones, modal ones, …
           provenance  //checks that the sources (agents/files) are
                       // represented (LD:Attribution) and the creation dates
                       // too (LD:History); LD:Authoritative is for checking
                       // if the author is a credible authority in the domain
     loss-less_integration  //checks that the semantics of
                            // source objects was not changed; the data model of
                            // WebKB and its protocols permits such an integration
     PD:timeliness  //alias SF:timeliness and LD:Currency; is
                    // the object is up-to-date? E.g., PD:newness (timely
                    // creation?) and PD:freshness (timely update?)
     SF:licensing  //alias, LD:licensed; to check for an open
                   // license, use PD:openness
     security  //checks for signatures, encryption,
               // maintainability (LD:sustainable), …
```

Figure 6. Important functions to evaluate the quality of the representation of terms

## 3  Description medium quality

Description-medium quality functions evaluate the textual/graphic/... presentation that some (kinds of) description-medium objects permit for some (kinds of) description content objects in some (kinds of) description containers. The more the "presentation is distinguished from content" and the more structured (fine-grained) the content, the more the presentation can be finely adapted for different kinds of users and by the end-users. To that end, the W3C advocates the use of XML-based languages (e.g., RDF/XML) as well as CSS, XSLT and GRDDL. This last language indicates which XSLT scripts can be used to translate some knowledge published in some XML-based KRLs into other knowledge representation languages (KRLs). Fresnel [28] may be seen as a kind of advanced CSS for RDF-based KRLs.

Presentation evaluation functions may for example give high values to graphical/textual/audio/... interfaces composed of fine-grained objects with "rich contextual menus". For each object, such menus would list i) presentation attributes/commands for this object, and ii) semantic relations/commands from/to/about this object to ease navigation, querying or updates. Although a syntax is clearly a presentation object, neither XML nor any current KRL seem to allow people to adapt their syntaxes, e.g., via the setting of some variables or the use of a notation ontology. Yet, this approach would be more flexible and easier to use than GRDDL, and hence can be used as one criteria by description-medium quality evaluating functions. Figure 7 gives a top-level specialization hierarchy for such functions. The general comments on the previous hierarchies also apply here: conventions, abbreviations, rationale for the specialization relations, etc.

```
description-medium_quality   //subtype of function type pm:quality
  quality_of_this_descr_medium   //to evaluate the source
        // object on all its criteria (→ aggregations of measures)
    descr_medium_quality_of_this_descr_medium
    quality_of_the_descr_content_related_to_this_descr_medium
    quality_of_the_descr_containers_related_to_descr_medium
  use_of_standard_formats   //for used KRLs (RDF/XML, …),
      // for character encodings, graphics (SVG, …), … (see w3.org);
      //3rd basic rule of Linked Data but for concrete models only
    use_of_structured_formats   //e.g., an HTML presentation with or
                                 // without RDFa statements
      use_of_formats_distinguishing_structure_from_presentation
          //e.g., XML but note that XML does not permit its users
          // to adapt its notation via the setting of some values
        use_of_notations_that_can_be_adapted_by_the_user
            // unlike XML and almost all notations
    use_of_machine-understandable-formats
      use_of_formats_that_have_an_interpretation_in_some_logic
        PD:format_interpretability   //aggregation of measures on
                       // qualities of formats proposed by a KB
          PD:human_and_machine_interpretability   //e.g., N3 can
                       // be more easily read than RDF/XML
  format_structural_quality   //subtypes on the next column
  format_concision   //e.g., N3 is more concise than RDF/XML
  format_uniformity   //reports on the extent to which similar
                       // things can be (re)presented in similar ways (from
                       // a software viewpoint and/or from a person viewpoint)
    SF:Uniformity   //pm:format_uniformity for a whole KB
  performance_of_this_format_for_this_task
     (description_medium, task -> value)   //function signature
```

```
/* Figure 7 continues here to detail the following subtype branch */
  format_structural_quality   //see the previous column
    format_abstract-expressiveness   //the expressiveness of
        // its abstract model (→ first-order_logic, …, kinds of
        // possible quantification (note: KIF allows to define
        // all kinds of relations to represent numerical quantifiers
        // but has no predefined keywords for them; thus, numerical
        // quantifiers defined by different users will be hard to match
        // (especially via simple graph-matching based techniques);
        // hence, KIF is expressive but low-level
    syntactic_expressiveness   //the higher the numeric result of
        //this function, the higher-level the notation can be considered
        // (for the selected criteria), i.e., the more flexible and
        // readable the format is, the more normalized/uniform
        // the descriptions are, and hence the easier to compare
        // via graph-matching these descriptions are
      syntactic_constructs_for_logical_constructs   //e.g., does
          // the format include keywords for numerical quantifiers
          // (e.g., "58%", "2 to 6") and for which kinds of them
      syntactic_constructs_for_creating_shortcuts   //kinds
          // of lambda-abstractions, …
      syntactic_constructs_for_ontological_primitives
         //e.g., for type partitions and/primitives such as those in
         // Ontoclean and extensions of them. They are needed
         // for knowledge engineering [3]. RDF is low-level: it
         // has no keywords for them but can import a
         // language ontology which has them
    referable_first-order-entities   //e.g., what can be a
        // 1st-order entity, i.e., what can be referred to via a
        // variable in the notation: concept nodes, relation nodes,
        // quantifiers, …; the more things can be 1st-order entities
        // (and hence that can be related to other things,
        // annotated, selected via a mouse, …), the better, and
        // the more formally related, the better for structuring or
        // annotation flexibility purposes;
        // from that viewpoint, an interface or notation for a KB
        // may be better than one for a database or a
        // structured document (which is then also better than an
        // unstructured one)
```

Figure 7. Important functions to evaluate the quality of a description medium

## 4  Description container quality

Description-containers quality functions evaluate the way a given description container – e.g., a static file or a distributed KB server – i) modularizes, stores, makes retrievable and accessible (i.e., how it "publishes") description content objects, and ii) checks or allows updates or queries on these objects. Compared to the independent and direct use of static files (e.g., RDF files), the use of knowledge servers eases knowledge modeling and reduces the implicit inconsistencies and redundancies between their knowledge statements. A KB server can also use static input/output files and offers much more flexibility than static files. It can also provide more services than those of a description-container (e.g., it can forward queries). This can be taken into account for evaluating its quality. Figure 8 gives a specialization hierarchy for description-container quality evaluating functions. The general comments on the previous hierarchies still apply.

```
description-container_quality   //subtype of function pm:quality
  quality_of_this_descr_container
    descr_container_quality_of_this_descr_container
    quality_of_the_descr_content_related_to_this_descr_container
    quality_of_the_descr_media_related_to_descr_container
    quality_of_the_processes_supported_by_this_descr_container
  storage_related_quality
    maximal_size_of_the_KB
    container_based_modularization
      static_container_based_modularization   //static file based
      dynamic_container_based_modularization   //forwarding or
              // replication of knowledge/queries amongst KBs
      LD:connectedness   //do combined datasets join correctly?
  assertion_related_quality   //what can be added or updated, by
                              // whom, in which language, …
    ontological_flexibility   //is the ontology fixed, i.e., is the
                              // KB actually just a database?
    LDpattern:annotation   //are third-party resources accepted?
    LDpattern:progressive_enrichment   //ways data (model) can
                                       //be improved over time
    checking_possibilities   //what kinds of inconsistencies or
                             // redundancies or redundancies can be detected?
                             //does the server advocate best practices to its users?
  information_retrieval_related_quality   //on the whole KB or on
                                          // some of its statements
    published_or_given_metadata   //on the KB or a part of it, e.g.,
         //via a "topic" (Ldpattern:Document_Type), via the use of a
         // semantic sitemap [11], voiD or DCAT, via metadata given
         // for any object (if a user requests it) but calculated in a
         // predefined way (as with "Concise Bounded Descriptions")
         // [2], or via metadata accessible via powerful queries
    object_accessibility
      PD:accessibility   //access methods, e.g., via SPARQL, an API,
               //                                a file (HTML,RDF)
      PD:availability   //percentage of time a given service is "up"
      SF:performance   //low latency, high throughput, only minor
                       // "performance variations", …
      PD:response_time   //e.g., for static access, SPARQL access
      PD:robustness   //average of performance over time; helped
                      // data cache and the use of LDpattern:parallel_loading
      querying_possibilities   //what can be queried, with which
                               // input/ouput languages, what privacy techniques are
                               // used, are the results ranked, filtered and merged, …
  interface_personalization   //to which extent can the input/output
    // presentation be adapted by end-users and can take into account
    // their constraints: language, disabilities,
    //                   access from various devices (mobile ones, …),
    //                   access from various software (browsers, …), …
```

Figure 8. Important ways to evaluate a description medium structural quality

## 5 Some other categorizations

In order to show how this knowledge criteria/quality ontology extends, generalizes and organizes the elements of its sources, Figure 9 lists show the structure of the four most organized sources that so far seemed to exist for Linked Data, even though they are essentially only two level deep. "SF:" is for [21], "Kahn" is for [19], "LDpattern:" is for [18] and "OPD:" is for [17] (this last source has three 3-level deep categories and one 4-level deep category). The first two sources are about quality criteria, the last two are about best practices. Their categories – the ones shown below – *seem to be* concept types. To permit a maximal integration of the various sources, they have been integrated into this quality ontology via function types, as illustrated by the previous indented lists. From these functions hierarchies, the concept types hierarchies can be generated. In the following lists of Figure 9, the lowermost categories are given within comments *and* without prefix for their source. The lowermost "OPD" subtypes have several instances in the OPD library.

```
SF:Quality_criterion   //this categorization often (but not always) follows
                      // the distinction between description content/medium/container
  SF:Content   //Consistency, Timeliness, Verifiability
  SF:Representation   //Uniformity, Versatility, Comprehensibility;
                      //mixes criteria on descr. medium and descr. container
  SF:Usage   //Validity of documents, Amount of Data, Licencing;
             // mixes criteria on descr. content and descr. container
  SF:System   //Accessibility, Performance

Kahn:Quality_dimension   //claimed to be "the 15 most important
                          //    ones from consumer perspective"
  Kahn:Intrinsic   //Believability, Accuracy, Objectivity, Reputation
  Kahn:Contextual   //Value-added, Relevancy, Timeliness,
                    // Completeness, Appropriate amount
  Kahn:Representational   //Interpretability, Ease of understanding,
                          // Consistency, Concision
  Kahn:Accessibility   //Accessibility, Access security

LDpattern:Linked_Data_pattern
  LDpattern:Identifier_pattern   //Hierarchical URIs, Literal Keys,
                 // Natural Keys, Patterned URIs, Proxy URIs,
                 // Shared Keys , URL Slug
  LDpattern:Modelling_pattern   //Custom Datatype, Index Resources,
          // Label Everything, Link Not Label, Multi-Lingual Literal,
          // N-Ary Relation , Ordered List, Ordering Relation,
          // Preferred Label , Qualified Relation, Reified Statement,
          // Repeated Property, Topic Relation, Typed Literal
  LDpattern:Publishing_pattern   //Annotation, Autodiscovery,
           // Document Type, Edit Trail, Embedded Metadata,
           // Equivalence Links, Link Base, Materialize Inferences,
           // Named Graphs, Primary Topic, Autodiscovery,
           // Progressive Enrichment, SeeAlso
  LDpattern:Application_pattern   //Assertion Query, Blackboard,
         // Bounded Description, Composite Descriptions,
         // Follow Your Nose, Missing Isn't Broken, Parallel Loading,
         // Parallel Retrieval, Resource Caching, Schema Annotation,
         // Smushing, Transformation Query

ODP:Ontology_Design_Pattern
  ODP:Structural ODP   //Architectural ODP, ODP:Logical ODP
    ODP:Logical ODP   //Logical_macro_ODP, Transformation_ODP
  ODP:Correspondence ODP
    ODP:Alignment ODP
    ODP:Re-engineering ODP
      ODP:Schema_reengineering_ ODP   //Refactoring_ ODP
  ODP:Content ODP, ODP:Reasoning ODP, ODP:Lexico-syntactic ODP
  ODP:Presentation ODP   //Naming ODP, Annotation ODP
```

Figure 9. Other categorizations for some elements in the quality ontology

# 6   Conclusions

This article has presented the top-level of an ontology organizing knowledge sharing best practices, design patterns, evaluation criteria and evaluation measures in a systematic, non-redundant and scalable way (e.g., by being based on distinctions on information objects rather than on processes). Some other research works on this subject mainly proposed *lists* of categories with, sometimes, some implementations (e.g., via SPARQL). The integration of these categories into this quality ontology shows that the results of these works cover only particular cases, which could sometimes be easily generalized. This ontology also permits to have a more synthetic view of the *kinds* of things that could or should be evaluated or done during knowledge sharing, or proposed by knowledge engineering tools. This ontology can be extended by Web users via the server which hosts it [29]. It could then be used as an index for elements of other libraries or ontologies. To that end, the bigger it will become, the more useful it will be. The presented ontology is, in some senses, validated by the fact it includes – or can be specialized to include – any quality related measures or criteria that the author has come across and by the fact it follows the strongest best practices it categorizes. Such an ontology is clearly application-independent. No particular use case would further validate it.

Section 2 also quickly introduced the data model of WebKB which enables i) loss integration of knowledge from various sources, and ii) the use of "default beliefs/rules/measures" to allow the combination of simple evaluation functions into complex ones and the re-use of other agents' functions. The fact that knowledge on the Semantic Web is full of implicit contradictions and redundancies, very hard to evaluate, and often incorrect even with respect to the OWL primitives that it re-uses [27], may be an indication that KBMS developers and knowledge providers to the Semantic Web would benefit from such a data model and such an ontology of best practices and quality measures.

# 7   References

[1] A. Palma, P. Haase, Y. Wang, M. d'Aquin. "D1.3.1 propagation models and strategies". NeOn deliv. D1.3.1, 2007.

[2] T. Heath, C. Bizer. "Linked Data: evolving the Web into a global data space". Synthesis Lectures on the Semantic Web: Theory and Technology, 1:1, 1–136. Morgan&Claypool, 2011.

[3] G. Guizzardi, M. Lopes, F. Baião, R. Falbo. "On the importance of truly ontological representation languages". IJISMD 2010. ISSN: 1947-8186.

[4] P.F. Patel-Schneider, "A revised architecture for Semantic Web reasoning". PPSWR 2005 (Germany), LNCS 3703, 32–36.

[5] M.R. Genesereth, R.E. Fikes, "Knowledge Interchange Format". Version 3.0. Technical Report Logic-92-1, Stanford Uni., 1992.

[6] P. Hayes, C. Menzel, J. Sowa, T. Tammet, M. Altheim, H. Delugach, M. Gruninger. "Common Logic (CL): a framework for a family of logic-based languages". ISO/IEC IS 24707:2007.

[7] A. Farquhar, R. Fikes, J. Rice. "The Ontolingua server: a tool for collaborative ontology construction". International Journal of Human-Computer Studies, 1996.

[8] S. Borgo, C. Masolo. "Ontological foundations of DOLCE". Handbook on Ontologies, Springer, 361–382, 2009.

[9] N. Guarino, C. Welty. "Evaluating ontological decisions with OntoClean". Comm. of the ACM, vol. 45(2), 61–65, 2002.

[10] ISWC 2006. "OWL specification of the Semantic Web (SW) Topics Ontology". http://lsdis.cs.uga.edu/library/resources/ontologies/swtopics.owl

[11] R. Cyganiak, R. Delbru, H. Stenzhorn, G. Tummarello, S. Decker. "Semantic sitemaps: efficient and flexible access to datasets on the semantic web". *ESWC 2008* (Tenerife, Spain), LNCS 5021, 690–704.

[12] K. Alexander, R. Cyganiak, M. Hausenblas, J. Zhao. "Describing linked datasets". *LDOW 2009* (Madrid, Spain).

[13] F. Maali, J. Erikson, P. Archer. "DCAT catalog vocabulary". W3C Working Draft, 2012.

[14] J. Breuker, W. van de Velde. "CommonKADS library for expertise modeling: Reusable Problem Solving Components". IOS Press, 1994.

[15] G. Dromey. "Scaleable formalization of imperfect knowledge". AWCVS 2006 (Macao), 21–33.

[16] J.Z. Pan, L. Lancieri, D. Maynard, F. Gandon, R. Cuel, A. Leger. "Success stories and best practices". Deliverable D1.4.2v2 of KWEB (Knowledge Web), EU-IST-2004-507482.

[17] V. Presutti, A. Gangemi. "Content Ontology Design Patterns as practical building blocks for web ontologies". ER 2008 (Spain) http://ontologydesignpatterns.org.

[18] L. Dodds, I. Davis. "Linked Data patterns – a pattern catalogue for modelling, publishing, and consuming Linked Data". http://patterns.dataincubator.org/book/, 56 pages, 2011.

[19] B.K. Kahn, D.M. Strong, R.Y. Wang. "Information quality benchmarks: product and service performance". Communications of the ACM, vol. 45(4) 2002, 184–192.

[20] G. Mcdonald. "Quality indicators for Linked Data datasets". http://answers.semanticweb.com/questions/1072/quality-indicators-for-linked-data-datasets (2011).

[21] A. Flemming, O. Hartig. "Quality criteria for Linked Data sources". http://sourceforge.net/apps/mediawiki/trdf/index.php?title=Quality_Criteria_for_Linked_Data_sources (2010).

[22] P.N. Mendes, C. Bizer, Y.H. Young, Z. Miklos, J.P. Calbimonte, A. Moraru. "Conceptual model and best practices for high-quality metadata". Deliverable 2.1 of PlanetData, FP7 project 257641 (2012).

[23] C. Bizer. "Quality-driven information filtering in the context of web-based information systems". PhD dissertation (195 pages), Free University of Berlin, 2007.

[24] C. Mader. "Quality criteria for SKOS vocabularies". https://github.com/cmader/qSKOS/wiki/Quality-Criteria-for-SKOS-Vocabularies (2012).

[25] C. Fürber. "Data quality constraints library". http://semwebquality.org/documentation/primer/20101124/ (2010).

[26] A. Gómez-Pérez, F. Ciravegna. "SEALS EU infrastructures project – semantic tool benchmarking". http://www.seals-project.eu/ (2012).

[27] A. Hogan, A. Harth, A. Passant, S. Decker, A. Polleres. "Weaving the pedantic web". LDOW 2010 (Raleigh, USA).

[28] E. Pietriga, C. Bizer, D. Karger, R. Lee. "Fresnel: a browser-independent presentation vocabulary for RDF". ISWC 2006 (USA), LNCS 4273, 158–171.

[29] Ph. Martin. "Collaborative knowledge sharing and editing". IJCSIS vol. 6(1), 2011, 14–29. ISSN: 1646-3692.